\documentclass[aps,twocolumn]{revtex4}

\usepackage{amssymb,graphicx}

\begin{document}

\title{Reply to Comment on ``Exposed-key weakness of $\alpha \eta$''}
\author{Charlene Ahn}
\email{cahn@toyon.com}
\affiliation{
Toyon Research Corporation,
6800 Cortona Drive,
Goleta, CA 93117}
\author{Kevin Birnbaum}
\email{Kevin.M.Birnbaum@jpl.nasa.gov}
\affiliation{
JPL M/S 161-135,
4800 Oak Grove Drive, Pasadena, CA 91109
}

\begin{abstract}
We address criticism of the Letter ``Exposed-Key Weakness of $\alpha \eta$'' in the Comment by Nair and Yuen. The Comment claims that the Letter does not show insecurity of $\alpha \eta$ because our approximation for the eavesdropper's entropy on the encrypted key is invalid. We present simulations which show that, on the contrary, our estimate is in close agreement with numerical calculations of the actual entropy over the applicable domain. We additionally discuss some ways in which our views on security requirements differ from the views given in the Comment.
\end{abstract}

\maketitle

The Comment by Nair and Yuen \cite{comment} claims to refute statements made in our Letter ``Exposed-Key Weakness of $\alpha\eta$'' \cite{letter}.  The main dispute concerns the validity of an approximation used in the derivation of our estimate for the eavesdropper's entropy on the encryption key. In this Reply, we present simulations supporting the validity of our estimate and provide a more detailed explanation of the reasoning underlying our approximation. We then give various applications of our estimate in quantifying security, as well as discuss some ways in which we differ from the authors of the Comment with regard to security requirements.

First, let us reiterate the claims made in the original Letter which are nominally disputed.  We assume an $\alpha\eta$ system using $M$ coherent states, initialized with $L$-bit seed key $K$, with measurement error described by a gaussian distribution with standard deviation $\sigma$. In our Letter, we state that even if an eavesdropper Eve starts with zero information on both the key and the message under transmission (i.e. a ciphertext only attack), 
\begin{quote}
We therefore take $U$\ldots as an upper bound on Eve's information on K per measured symbol.  We expect Eve's information to grow linearly with the number of symbols\ldots This approximation will of course break down when the Eve's entropy on $K$ is low, such that her entropy on the key will only asymptotically approach zero as the number of symbols goes to infinity\ldots Eve's entropy on the key will transition from linear decline to asymptotic decay after measuring approximately $n_0 = L/U$ symbols\ldots
\end{quote}
where we derived $U$ to be
\begin{equation}
U \approx \log(\frac{M}{\sigma \sqrt{2 \pi e}}) - 1.
\end{equation}
This can be summarized as 
\begin{equation}
H_E(K) \geq L - QU,
\end{equation}
\begin{equation}
\label{initial}
H_E(K) \approx L - QU \quad \mathrm{for} \quad Q\ll n_0 = L/U,
\end{equation}
where $Q$ is the number of encoded bits sent, and
\begin{equation}
\label{asymptotic}
\lim_{Q\to \infty} H_E(K) = 0,
\end{equation}
where $H_E(K)$ is Eve's entropy on the key.

The authors of the Comment argue that (\ref{initial}) can be replaced with 
\begin{eqnarray}
\label{equality}
H_E(K) = L-QU \quad \mathrm{for} \quad Q\leq n_{dep}\\
\label{bound}
H_E(K) \geq L-QU \quad \mathrm{for} \quad Q > n_{dep},
\end{eqnarray}
where $n_{dep}$ is the number of statistically independent strings $\{k_q\}$ generated by the pseudo-random number generator (PRNG), and is dependent on which PRNG is used; an upper bound for $n_{dep}$ is given in the Comment of $n_{dep} \leq |K|/\log_2 M/2$.  We do not disagree with (\ref{equality},\ref{bound}).

 However, the Comment additionally claims that for $n_{dep} < Q \ll n_0$, the left and right hand sides of (\ref{bound}) may be sufficiently far apart such that the approximation (\ref{initial}) is not valid.  To investigate this claim, we performed Monte Carlo simulations of the $\alpha \eta$ system.  For each simulation, a seed key and message were chosen from a uniform distribution. The running key was generated from the seed key using an $L$-bit linear feedback shift register (LFSR) as the PRNG. Eve's measurements were simulated by adding a gaussian distributed random variable to the phase angle of each symbol sent.  We calculated the probability that Eve assigns to each seed key by starting with a uniform probability and using the measurement result after each transmitted symbol to update the probabilities.  

\begin{figure}
   \begin{center}
   \includegraphics[width=0.5\textwidth]{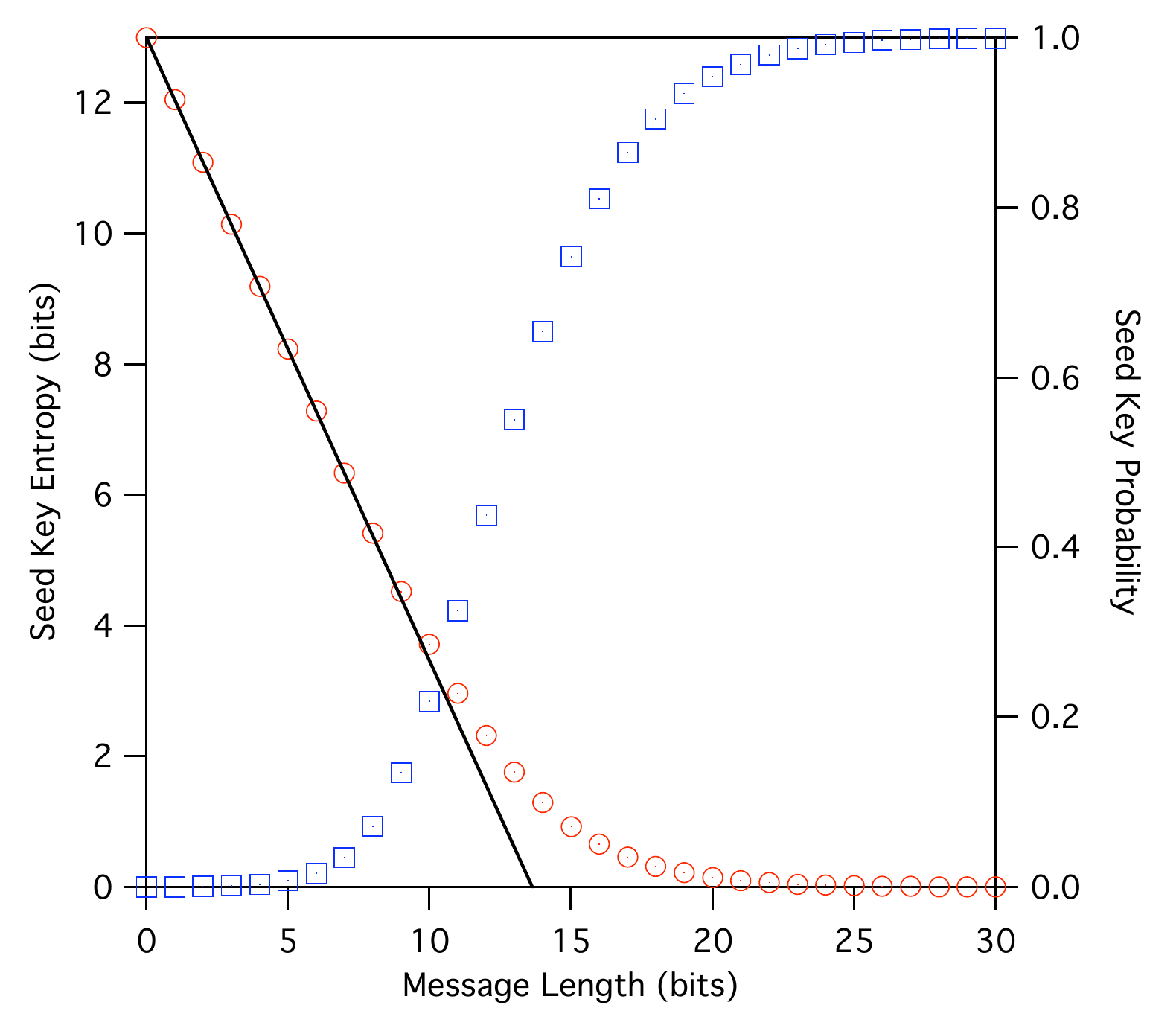} 
   \end{center}
   \caption{Circles ($\odot$) on the left axis denote the expectation of Eve's entropy on the key, $H_E(K)$, as a function of the number of encoded bits sent, $Q$.  Squares ($\boxdot$) on the right axis indicate $P_E(K)$ the average probability that Eve assigns to the correct value of the key.  The line is the estimate (\ref{initial}) calculated in the Letter.  The PRNG is a 13-bit LFSR, the number of possible symbols for each transmission is $M=256$, and the gaussian measurement noise has standard deviation $\sigma = 16$.  Averaging was performed over $10^4$ Monte-Carlo simulations.}
   \label{fig:monte-carlo}
\end{figure}

Over this ensemble, with randomly chosen seed, message, and measurement noise, we computed the average entropy that Eve has on the seed key, and the average probability that she assigns to the correct seed key, $P_E(K)$ (see Figure~\ref{fig:monte-carlo}).  The parameters used in the calculation are $L=13$, $M=256$, $\sigma=16$, and the averaging was performed over $10^4$ simulations.  For comparison, our estimate (\ref{initial}) is also plotted.

As can be seen from Fig.~\ref{fig:monte-carlo}, our estimate is quite accurate over the specified domain.  Note that with these simulation parameters, $n_{dep} < 2$; that is, the second running key is not statistically independent of the first.  However, it can be seen that the amount of information that Eve gains from each symbol is nearly the same for the first several symbols.  Indeed, even when the entropy is less than the size of the running key, our estimate is still valid.  The actual entropy does not significantly deviate from our estimate until the entropy is quite low.

This can be understood by re-visiting the derivation in the Letter.  The derivation does not depend on an analogy to Shannon's random cipher, as the Comment supposes, and additionally is completely independent of the contents of the plaintext.  It instead results from a consideration of the probabilities that Eve assigns to the keys after each measurement.  

Eve begins with no information on the key.  Therefore, she assigns the same probability to all possible values.  Upon making a measurement, however, she can update these probabilities according to Bayes' rule, which states that the probability of each key should be multiplied by the probability that that key would generate the observed measurement (divided by a normalization factor).  Thus seed keys which generate running keys close (on the half-circle) to the observed symbol will have their probabilities increased, while seed keys which generate distant running keys will have their probabilities lowered. The information that Eve gains is related to the change in the probabilities she assigns to the symbols before and after the measurement.


The estimate (\ref{initial}) is based on the approximation that just prior to each measurement, Eve's probabilities are, on average, nearly uniform across the possible symbol values, and assumes a well-behaved PRNG.
We may consider the PRNG as an ordered list of $Q$ maps from the space of seed keys $\{0,\dots 2^L-1\}$ to the space of running keys $\{0,\dots M-1\}$.  We assume that the PRNG will be a typical member of the set of all lists of maps, and therefore that the distribution of running keys (on ensembles with varying seed key and iteration number) will be nearly uniform.  In other words, though the PRNG is entirely deterministic, its ensemble distributions will mimic those of a true random number generator with uniform probabilities.  In practice, even an LFSR, which is generally considered a poor PRNG, appears to be sufficiently well-distributed for our purposes.

Let us consider a simple case in order to better understand why this estimate works.  Take the case where the measurement noise is uniformly distributed over one quarter of the phase circle.  The determination of the half-circle encodes the data.  The knowledge of the quarter circle may be used to reject, after each measurement, the possibility of one half of the possible running keys.  Approximately one half of the seed keys will generate one of the running keys that can be ruled out.  Therefore those seed keys can be eliminated.  Since the running keys for different seeds are not correlated, we are not throwing out the exact same set of seed keys over and over; after each measurement, we are throwing out one half of the seed keys, selected in a uniform fashion.  At any step, about as many remaining seed keys generate running keys within the correct quadrant as outside of it; thus we can eliminate approximately half of the \emph{remaining} keys after each measurement.  This is true even when $Q>n_{dep}$; that is, statistical independence is not required.

This estimate does not break down until the estimated number of remaining seed keys is of order one.  It breaks down because there remains the possibility that there will be another seed key which will generate running keys which, like the running keys generated by the correct seed key, are all in the same quadrant as the measurements.  The probability that this is true for any given false seed key, however, falls exponentially with the number of transmission events.  This ``probability'' involves both the truly random measurement noise and the frequency of occurrence of events in an ensemble generated by the deterministic PRNG.

In summation, the estimate (\ref{initial}) is based on an approximation which continues to hold until Eve's entropy is low (of order one bit) or, equivalently, the probability she assigns to the correct key is high (of order one).  It does not rely on the statistical independence of the running keys.

In the region about $n_0$, the behavior of the entropy function becomes more complicated.  At the present time, we do not have a precise analytic form for the entropy in this region, and so we are limited to numerical simulations such as those shown above.  The authors of the Comment claim that no statements about the insecurity of the system may be made until a precise expression, applicable over the entire domain $Q\in[0,\infty)$, is found for the probability that Eve assigns to the correct seed key.  While we agree that such an expression would be desirable, we believe that useful statements about the security of $\alpha\eta$ can be made even without this expression.

For example, consider a user who deploys an $\alpha\eta$ system with the parameters in the simulation above.  She may specify a security requirement to meet the needs of her application.  For example, she may require that the eavesdropper's entropy on the key must be more than $5$ bits under a ciphertext-only attack.  For this system under this requirement, it is clear from Fig.~\ref{fig:monte-carlo} that the system is insecure (i.e. does not meet the requirement) for message lengths above $8$ bits.  This is a well-quantified statement.  It is also clear that she could accurately make this statement based on the estimate (\ref{initial}).


Requiring a limit on Eve's entropy as we do above is a common measure of information-theoretic security. For example, in quantum key distribution experiments, a typical requirement is for Eve's expected information on the key to be less than $10^{-6}$ bits for a secret key of hundreds of bits \cite{hughes:free-space-QKD}.  The authors of the Comment seem to make the implicit assumption that all security requirements will be stated in the form of a maximum probability that Eve will assign to the correct key. However, even when limited to this sort of security requirement, statements about the security can still be made. 

We note that $P_E(K)\geq 2^{-H_E(K)}$.  Therefore, knowledge of $H_E(K)$ can be used to show that a maximum bound on $P_E(K)$ is violated, though it is not sufficient to determine that the bound is obeyed.  For example, consider an $\alpha\eta$ system with the parameters above, with a security requirement of $P_E(K)<2^{-5}$.  As seen in Fig.~\ref{fig:monte-carlo}, the user could accurately determine from the estimate of $H_E(K)$ that the system would not meet the requirement for message lengths of 9 or more bits.  In fact, the system would also be insecure for message lengths of 7 or 8 bits, so she could not use the estimate to find the region of security, but she could determine some message lengths as insecure.

If the security requirement of the user involves an entropy of order one bit or less, or a probability of order one, then the transition will be outside the domain of our estimate (\ref{initial}).  In principle, therefore, we would require additional analysis to meet the needs of all users.  In practice, however, user requirements tend to be much more stringent.  For example, the experiments in \cite{hughes:free-space-QKD} typically sacrifice about half the bits for additional privacy amplification so that Eve's knowledge of a fraction of order one bits of the secret key is reduced to the previously mentioned $10^{-6}$ bits of information on hundreds of bits of secret key. Thus, for $\alpha \eta$, we might expect typical user requirements to be from a few tens to a few hundreds of bits of entropy on the key.  In this region, the estimate (\ref{initial}) will be quite accurate, and should be adequate to find the maximum secure message length.

Another instance where our views on security requirements differ from those of the Comment's authors can be seen by considering the implications of the limit (\ref{asymptotic}).  This equation guarantees that for any security requirement with a non-zero key entropy (or a maximum key probability less than one), there exists a message length such that the system will be insecure under a ciphertext-only attack.  This was stated more informally in the Letter as ``\dots Eve may have enough information to determine the key with high probability when $Q\gg n_0$.''  The Comment claims that these statements are ``unfalsifiable'' and ``do not satisfy the requirement of being a scientific claim.''  However, even in the absence of an analysis such as that given by Figure \ref{fig:monte-carlo}, we believe that our statements are not trivial.  In the Letter, we provide a counter-example: the simple additive streaming cipher.  For this cipher, $H_E(K)=L$ for all values of $Q$ under the ciphertext-only attack.  Thus there exists no $Q$ such that Eve may determine the key with high probability (where ``high'' may be chosen by the user to have any value above $2^{-L}$).  By disproving our statement for another cipher, we prove that the statement itself is falsifiable.

In contrast, the authors of the Comment compare the additive stream cipher to $\alpha \eta$ by claiming that  ``Intuitively, the measurement noise in $\alpha\eta$ would make it more secure than an additive stream cipher instead of worse as claimed in [the Letter] at least for the case of known-plaintext attacks where $H(\mathbf{X}^n)=0$.''  This intuition rests on a particular selection of security requirement contrary to the ones we have discussed above: that is, $H_E(K)=0$.  While, technically speaking, the user may choose any security requirement, we again note that in practice users generally have more stringent requirements.  For example, we do not know of any application in which the eavesdropper may know the key with confidence $0.99$ and yet the system is considered secure.

This approach to security is perhaps related to the choice in the Comment to focus on the quantity $\bar{N}_k$, the number of false seed keys to which Eve assigns a non-zero probability.  As the Comment points out, the tail of the gaussian noise distribution does not reach zero at any point on the circle. Eve's seed key probabilities are the product of the conditional probabilities of the measurements. The conditional probabilities are all non-zero.  Any product of a finite number of non-zero numbers is also non-zero.  Hence for any finite $Q$, Eve will assign a non-zero probability to each seed key.  Therefore $\bar{N}_k = 2^L -1$ for all finite $Q$ (though $\lim_{Q\to \infty} \bar{N}_k = 0$).

Though we agree with this analysis, we do not believe it is relevant to users in practice.  If it were, the user could make the gaussian noise $\sigma$  arbitrarily small (say, $\sigma = 10^{-10}$).  The eavesdropper would correctly determine each running key and message bit transmitted, with a probability close to one, but could never drive the entropy exactly to zero, or completely rule out the highly improbable messages and keys.  But in this case the user could achieve the same result by simply transmitting the plaintext over a channel with an arbitrarily small (but non-zero) bit error rate.  Since in practice no channel has a bit error rate which is precisely zero, we believe that if this were the only security requirement the user had, then any real channel would suffice without the need for an encryption system such as $\alpha\eta$.

We close by responding to another criticism of our Letter. The authors of the Comment also hold that ``there is no commonly agreed meaning of the symbols `$\approx$' and `$\ll$'" and therefore our estimate (\ref{initial}) ``is not well-defined\dots it cannot be \emph{falsified}, the possibility of the latter being the hallmark of a meaningful scientific statement'' (emphasis in original).  While we believe that exact equalities are preferable to approximations, we do not agree that all use of approximation is unscientific.  For example, we feel no qualms about statements such as 
\begin{equation}
\label{smallangle}
\tan(x)\approx x  \quad \mathrm{for} \quad x\ll 1,
\end{equation}
which was also used in the Letter, without argument from the authors of the Comment, or
\begin{equation}
\label{Nk}
(2^{H(K)}-1)2^{-nD}\approx 2^{H(K)-nD}  \quad \mathrm{for} \quad 2^{H(K)}\gg 1,
\end{equation}
which was used in the Comment itself (though, to be fair, they used ``$\doteq$'' instead of ``$\approx$'', and did not specify the assumption used in the approximation).

In conclusion, we do not find that the Comment refutes the claims of our Letter.  We have performed additional simulations which show that our estimates of the eavesdropper's entropy are quite accurate in the specified domain.  We also find that some of the additional claims made in the Comment, while technically true, are not relevant in practice to users of the $\alpha\eta$ system.

\bibliography{replybib}

\end{document}